\pdfoutput=1
\documentclass[12pt,a4paper]{article}

\usepackage{ifthen} 
\newboolean{pdflatex}
\setboolean{pdflatex}{true} 

\newboolean{articletitles}
\setboolean{articletitles}{true} 

\newboolean{uprightparticles}
\setboolean{uprightparticles}{false} 

\def\paperauthors{LHCb collaboration} 
\def\paperasciititle{Measurement of transverse Lambda hyperon polarization in  pPb 5 TeV} 
\def\papertitle{Measurement of transverse $\Lz$ and $\Lbar $ 
hyperon polarization in  $p$Pb collisions at \sqsnn  = 5.02\tev} 
\def\paperkeywords{{High Energy Physics}, {LHCb}} 
\def\papercopyright{\the\year\ CERN for the benefit of the LHCb collaboration} 
\def\paperlicence{CC BY 4.0 licence}
\def\paperlicenceurl{https://creativecommons.org/licenses/by/4.0/}

\newif\ifEnableSectionTOCLinks
\EnableSectionTOCLinksfalse 

\usepackage[top=1in, bottom=1.25in, left=1in, right=1in]{geometry}

\usepackage{microtype}
\usepackage{lineno}  
\usepackage{xspace} 
\usepackage{caption} 

\usepackage{graphicx}  
\usepackage{color}
\usepackage{colortbl}
\graphicspath{{./figs/}} 

\usepackage{amsmath} 
\usepackage{amssymb}
\usepackage{amsfonts}
\usepackage{upgreek} 

\newcommand*\patchAmsMathEnvironmentForLineno[1]{%
\expandafter\let\csname old#1\expandafter\endcsname\csname #1\endcsname
\expandafter\let\csname oldend#1\expandafter\endcsname\csname
end#1\endcsname
 \renewenvironment{#1}%
   {\linenomath\csname old#1\endcsname}%
   {\csname oldend#1\endcsname\endlinenomath}%
}
\newcommand*\patchBothAmsMathEnvironmentsForLineno[1]{%
  \patchAmsMathEnvironmentForLineno{#1}%
  \patchAmsMathEnvironmentForLineno{#1*}%
}
\AtBeginDocument{%
\patchBothAmsMathEnvironmentsForLineno{equation}%
\patchBothAmsMathEnvironmentsForLineno{align}%
\patchBothAmsMathEnvironmentsForLineno{flalign}%
\patchBothAmsMathEnvironmentsForLineno{alignat}%
\patchBothAmsMathEnvironmentsForLineno{gather}%
\patchBothAmsMathEnvironmentsForLineno{multline}%
\patchBothAmsMathEnvironmentsForLineno{eqnarray}%
}

\usepackage[pdftex,
            pdfauthor={\paperauthors},
            pdftitle={\paperasciititle},
            pdfkeywords={\paperkeywords}]{hyperref}
\usepackage{hyperxmp}
\hypersetup{
    pdfcopyright={Copyright (C) \papercopyright},
    pdflicenseurl={\paperlicenceurl}
}

\usepackage[colorinlistoftodos,textsize=scriptsize]{todonotes}

\usepackage[bottom,flushmargin,hang,multiple]{footmisc}

\usepackage[all]{hypcap} 

\usepackage{xspace} 
\usepackage{upgreek}

\def\lhcb   {\mbox{LHCb}\xspace}

\def\belle  {\mbox{Belle}\xspace}

\def\MagUp {\mbox{\em Mag\kern -0.05em Up}\xspace}

\ifthenelse{\boolean{uprightparticles}}%
{
 
 \def\Pgamma      {\ensuremath{\upgamma}\xspace}

 \def\Pmu         {\ensuremath{\upmu}\xspace}                 
 \def\Pnu         {\ensuremath{\upnu}\xspace}                 
                  
 \def\Ppi         {\ensuremath{\uppi}\xspace}

 \def\Ppsi        {\ensuremath{\uppsi}\xspace}

 \def\PDelta      {\ensuremath{\Delta}\xspace}                 
 \def\PXi         {\ensuremath{\Xi}\xspace}                 
 \def\PLambda     {\ensuremath{\Lambda}\xspace}                 
 \def\PSigma      {\ensuremath{\Sigma}\xspace}                 
 \def\POmega      {\ensuremath{\Omega}\xspace}                 
 \def\PUpsilon    {\ensuremath{\Upsilon}\xspace}
 \let\oldPi\Pi
 \def\PPi         {\ensuremath{\oldPi}\xspace}

 \def\PB      {\ensuremath{\mathrm{B}}\xspace}                 
 \def\PD      {\ensuremath{\mathrm{D}}\xspace}                 
 \def\PJ      {\ensuremath{\mathrm{J}}\xspace}                 
 \def\PK      {\ensuremath{\mathrm{K}}\xspace}                 

 \def\Pe      {\ensuremath{\mathrm{e}}\xspace}                 
 \def\Pp      {\ensuremath{\mathrm{p}}\xspace}                 

 \def\Ps      {\ensuremath{\mathrm{s}}\xspace}

 \def\thebaroffset{0.0em}
}
{
 
 \def\Pgamma      {\ensuremath{\gamma}\xspace}

 \def\Pmu         {\ensuremath{\mu}\xspace}                 
 \def\Pnu         {\ensuremath{\nu}\xspace}                 
                  
 \def\Ppi         {\ensuremath{\pi}\xspace}

 \def\Ppsi        {\ensuremath{\psi}\xspace}                 
                  
 \mathchardef\PDelta="7101
 \mathchardef\PXi="7104
 \mathchardef\PLambda="7103
 \mathchardef\PSigma="7106
 \mathchardef\POmega="710A
 \mathchardef\PUpsilon="7107
 \mathchardef\PPi="7105
 \def\PB      {\ensuremath{B}\xspace}                 
 \def\PD      {\ensuremath{D}\xspace}                 
 \def\PJ      {\ensuremath{J}\xspace}                 
 \def\PK      {\ensuremath{K}\xspace}                 

 \def\Pe      {\ensuremath{e}\xspace}                 
 \def\Pp      {\ensuremath{p}\xspace}                 

 \def\Ps      {\ensuremath{s}\xspace}

 \def\thebaroffset{0.18em}
}
\newcommand{\offsetoverline}[2][\thebaroffset]{\kern #1\overline{\kern -#1 #2}}%

\makeatletter
\ifcase \@ptsize \relax
  \newcommand{\miniscule}{\@setfontsize\miniscule{4}{5}}
\or
  \newcommand{\miniscule}{\@setfontsize\miniscule{5}{6}}
\or
  \newcommand{\miniscule}{\@setfontsize\miniscule{5}{6}}
\fi
\makeatother

\DeclareRobustCommand{\optbar}[1]{\shortstack{{\miniscule (\rule[.5ex]{1.25em}{.18mm})}
  \\ [-.7ex] $#1$}}

\def\en         {{\ensuremath{\Pe^-}}\xspace}   
\def\ep         {{\ensuremath{\Pe^+}}\xspace}

\def\mup        {{\ensuremath{\Pmu^+}}\xspace}
\def\mun        {{\ensuremath{\Pmu^-}}\xspace} 

\def\neu        {{\ensuremath{\Pnu}}\xspace}

\def\neum       {{\ensuremath{\neu_\mu}}\xspace}

\def\g      {{\ensuremath{\Pgamma}}\xspace}

\def\squark    {{\ensuremath{\Ps}}\xspace}

\def\pion   {{\ensuremath{\Ppi}}\xspace}

\def\pim    {{\ensuremath{\pion^-}}\xspace}
\def\pipm   {{\ensuremath{\pion^\pm}}\xspace}

\def\kaon    {{\ensuremath{\PK}}\xspace}

\def\KorKbar {\kern \thebaroffset\optbar{\kern -\thebaroffset \PK}{}\xspace}

\def\Kpm     {{\ensuremath{\kaon^\pm}}\xspace}

\def\KS      {{\ensuremath{\kaon^0_{\mathrm{S}}}}\xspace}

\def\D       {{\ensuremath{\PD}}\xspace}

\def\DorDbar {\kern \thebaroffset\optbar{\kern -\thebaroffset \PD}\xspace}

\def\Dp      {{\ensuremath{\D^+}}\xspace}
\def\Dm      {{\ensuremath{\D^-}}\xspace}

\def\DpDm    {\ensuremath{\Dp {\kern -0.16em \Dm}}\xspace}

\def\B       {{\ensuremath{\PB}}\xspace}

\def\BorBbar {\kern \thebaroffset\optbar{\kern -\thebaroffset \PB}\xspace}

\def\Bd      {{\ensuremath{\B^0}}\xspace}

\def\BdorBdbar {\kern \thebaroffset\optbar{\kern -\thebaroffset \Bd}\xspace}

\def\Bs      {{\ensuremath{\B^0_\squark}}\xspace}

\def\BsorBsbar {\kern \thebaroffset\optbar{\kern -\thebaroffset \Bs}\xspace}

\def\jpsi     {{\ensuremath{{\PJ\mskip -3mu/\mskip -2mu\Ppsi}}}\xspace}

\def\Y#1S{\ensuremath{\PUpsilon{(#1S)}}\xspace}

\def\proton      {{\ensuremath{\Pp}}\xspace}

\def\Lz          {{\ensuremath{\PLambda}}\xspace}
\def\Lbar        {{\ensuremath{\offsetoverline{\PLambda}}}\xspace}
\def\LorLbar     {\kern \thebaroffset\optbar{\kern -\thebaroffset \PLambda}\xspace}

\def\Sigmares    {{\ensuremath{\PSigma}}\xspace}
\def\Sigmaz      {{\ensuremath{\Sigmares{}^0}}\xspace}
\def\Sigmap      {{\ensuremath{\Sigmares{}^+}}\xspace}
\def\Sigmam      {{\ensuremath{\Sigmares{}^-}}\xspace}
\def\Sigmaresbar {{\ensuremath{\offsetoverline{\Sigmares}}}\xspace}

\def\Sigmabarm   {{\ensuremath{\Sigmaresbar{}^-}}\xspace}
\def\Xires       {{\ensuremath{\PXi}}\xspace}
\def\Xiz         {{\ensuremath{\Xires^0}}\xspace}
\def\Xim         {{\ensuremath{\Xires^-}}\xspace}
\def\Xiresbar       {{\ensuremath{\offsetoverline{\Xires}}}\xspace}

\def\Xibarp      {{\ensuremath{\Xiresbar^+}}\xspace}
\def\Omegares    {{\ensuremath{\POmega}}\xspace}

\def\Omegam      {{\ensuremath{\Omegares^-}}\xspace}

\newcommand{\decay}[2]{\mbox{\ensuremath{#1\!\to #2}}\xspace} 

\def\to                 {\ensuremath{\rightarrow}\xspace}

\newcommand{\etot}{{\ensuremath{\varepsilon_{\mathrm{ tot}}}}\xspace}

\def\AT#1     {\ensuremath{A_{\mathrm{T}}^{#1}}\xspace}           

\def\C#1      {\ensuremath{\mathcal{C}_{#1}}\xspace}                       
\def\Cp#1     {\ensuremath{\mathcal{C}_{#1}^{'}}\xspace}                    
\def\Ceff#1   {\ensuremath{\mathcal{C}_{#1}^{\mathrm{(eff)}}}\xspace}        
\def\Cpeff#1  {\ensuremath{\mathcal{C}_{#1}^{'\mathrm{(eff)}}}\xspace}       
\def\Ope#1    {\ensuremath{\mathcal{O}_{#1}}\xspace}                       
\def\Opep#1   {\ensuremath{\mathcal{O}_{#1}^{'}}\xspace}                    

\newcommand{\aunit}[1]{\ensuremath{\text{\,#1}}}       

\newcommand{\tev}{\aunit{Te\kern -0.1em V}\xspace}
\newcommand{\gev}{\aunit{Ge\kern -0.1em V}\xspace}
\newcommand{\mev}{\aunit{Me\kern -0.1em V}\xspace}
\newcommand{\kev}{\aunit{ke\kern -0.1em V}\xspace}
\newcommand{\ev}{\aunit{e\kern -0.1em V}\xspace}
 
\newcommand{\mevc}{\ensuremath{\aunit{Me\kern -0.1em V\!/}c}\xspace}
\newcommand{\gevc}{\ensuremath{\aunit{Ge\kern -0.1em V\!/}c}\xspace}
\newcommand{\mevcc}{\ensuremath{\aunit{Me\kern -0.1em V\!/}c^2}\xspace}
\newcommand{\gevcc}{\ensuremath{\aunit{Ge\kern -0.1em V\!/}c^2}\xspace}

\def\mm   {\aunit{mm}\xspace}

\def\nm   {\aunit{nm}\xspace}

\def\nb {\aunit{nb}\xspace}
\def\invnb {\ensuremath{\nb^{-1}}\xspace}

\newcommand{\chisq}{\ensuremath{\chi^2}\xspace}

\newcommand{\chisqip}{\ensuremath{\chi^2_{\text{IP}}}\xspace}

\def\deriv {\ensuremath{\mathrm{d}}}

\def\gsim{{~\raise.15em\hbox{$>$}\kern-.85em
          \lower.35em\hbox{$\sim$}~}\xspace}
\def\lsim{{~\raise.15em\hbox{$<$}\kern-.85em
          \lower.35em\hbox{$\sim$}~}\xspace}

\def\sqsnn {\ensuremath{\protect\sqrt{s_{\scriptscriptstyle\text{NN}}}}\xspace}
\def\pt         {\ensuremath{p_{\mathrm{T}}}\xspace}
\def\xf         {\ensuremath{x_{\mathrm{F}}}\xspace}

\def\evtgen     {\mbox{\textsc{EvtGen}}\xspace}

\def\geant      {\mbox{\textsc{Geant4}}\xspace}

\def\tell1  {TELL1\xspace}
\def\ukl1   {UKL1\xspace}

\newcommand{\lhcborcid}[1]{\href{https://orcid.org/#1}{\hspace*{0.1em}\raisebox{-0.45ex}{\includegraphics[width=1em]{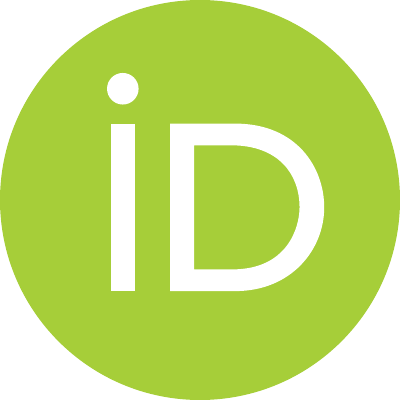}}}}

\hypersetup{
  colorlinks   = true, 
  urlcolor     = blue, 
  linkcolor    = blue, 
  citecolor    = red   
}

\ifEnableSectionTOCLinks
    \usepackage[explicit]{titlesec} 
    
    \let\oldcontentsline\contentsline
    \renewcommand

    \titleformat{\section}{\normalfont\Large\bf}{\hyperlink{tocsection.\thesection}{{\thesection} \parbox[t]{\dimexpr\textwidth-1pc}{#1}}}{1pc}{}

    \titleformat{\subsection}{\normalfont\bf}{\hyperlink{tocsubsection.\thesubsection}{{\thesubsection} \parbox[t]{\dimexpr\textwidth-1pc}{#1}}}{1pc}{}

    \titleformat{name=\section,numberless}[display]{}{}{0pt}{\normalfont\Huge\bfseries #1}
\fi

\usepackage{cite} 
\usepackage{LHCb/mciteplus}

\usepackage{longtable} 

\begin{document}

\renewcommand{\thefootnote}{\fnsymbol{footnote}}
\setcounter{footnote}{1}


\begin{titlepage}
\pagenumbering{roman}

\vspace*{-1.5cm}
\centerline{\large EUROPEAN ORGANIZATION FOR NUCLEAR RESEARCH (CERN)}
\vspace*{1.5cm}
\noindent
\begin{tabular*}{\linewidth}{lc@{\extracolsep{\fill}}r@{\extracolsep{0pt}}}
\ifthenelse{\boolean{pdflatex}}
{\vspace*{-1.5cm}\mbox{\!\!\!\includegraphics[width=.14\textwidth]{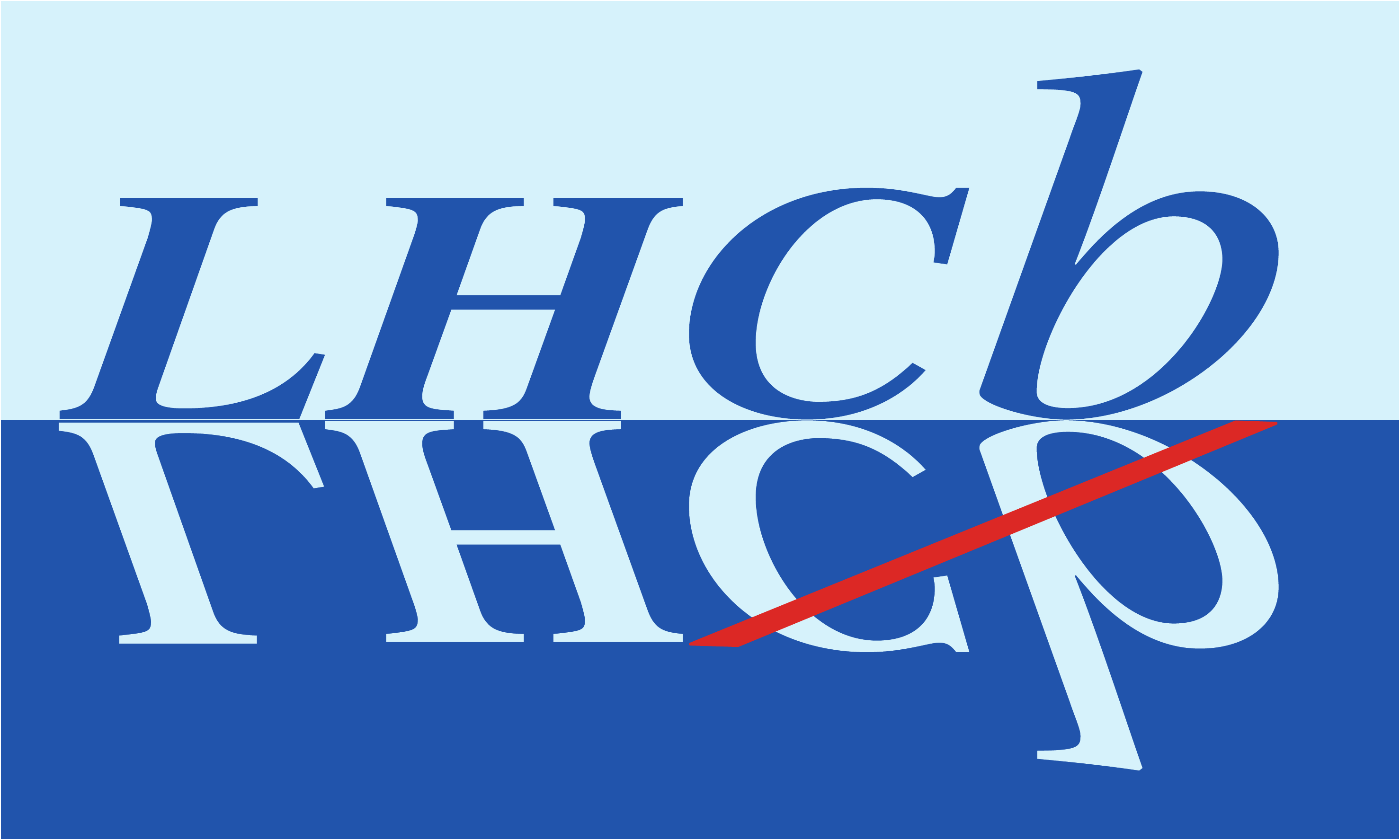}} & &}%
{\vspace*{-1.2cm}\mbox{\!\!\!\includegraphics[width=.12\textwidth]{LHCb/Figures/lhcb-logo.pdf}} & &}%
\\
 & & CERN-EP-2025-153 \\  
 & & LHCb-PAPER-2025-004 \\  
 & & December 24, 2025 \\ 
 & & \\
\end{tabular*}

\vspace*{4.0cm}

{\normalfont\bfseries\boldmath\huge
\begin{center}
  \papertitle 
\end{center}
}

\vspace*{2.0cm}

\begin{center}
\paperauthors\footnote{Authors are listed at the end of this paper.}
\end{center}

\vspace{\fill}

\begin{abstract}
  \noindent
The transverse polarization of $\Lz$ and $\Lbar$ hyperons is measured in $p$Pb collisions collected by the LHCb experiment at a nucleon-nucleon center-of-mass energy of $5.02 \tev$. The polarization is averaged over hyperon transverse momentum in the range $0.15 < \pt < 6.00 \gevc$, and Feynman-$x$ in the ranges $0.005 < \xf < 0.040$ (forward region) and $-0.10 < \xf < -0.01$ (backward region) defined relative to the proton beam direction. The transverse polarization is found to be compatible with zero for both $\Lz$ and $\Lbar$ hyperons. The results are also measured as a function of $\pt$ and \xf with no significant dependence on these variables observed. The results are compared with previous experimental measurements at different center-of-mass energies and collision environments. 
\end{abstract}

\vspace*{2.0cm}

\begin{center}
  Published in Phys.~Rev.~D112 (2025) 112022
\end{center}

\vspace{\fill}

{\footnotesize 
\centerline{\copyright~\papercopyright. \href{\paperlicenceurl}{\paperlicence}.}}
\vspace*{2mm}

\end{titlepage}


\newpage
\setcounter{page}{2}
\mbox{~}
%
%
%
%


\renewcommand{\thefootnote}{\arabic{footnote}}
\setcounter{footnote}{0}


\cleardoublepage


\pagestyle{plain} 
\setcounter{page}{1}
\pagenumbering{arabic}


\section{\label{sec:intro}Introduction}
Spontaneous transverse polarization in the production of $\Lz$ baryons, with values up to 30\%, was first measured nearly 50 years ago in unpolarized proton--proton ($pp$) and proton--nucleus collisions \cite{Bunce:1976yb, Heller:1978ty}, which contradicted early leading-order perturbative QCD calculations ~\cite{Kane:1978nd}, implying that the polarization must be due to nonperturbative aspects of QCD interactions. Since the first observation of transverse $\Lz$ polarization in hadron-hadron collisions, numerous experiments have explored this polarization using various beams and targets at different energies \cite{Schachinger:1978qs, Erhan:1979xm,Heller:1983ia, Lundberg:1989hw, Yuldashev:1990az, Ramberg:1994tk, Fanti:1998px, HERA-B:2006rds,ATLAS:2014ona}. The observed polarization is negative, shows little dependence on the beam energy, and increases with Feynman-$x$, $\xf = 2p_z^*/\sqsnn$, where $p_z^*$ is the longitudinal momentum of the hyperon in the center-of-mass frame, and transverse momentum, \pt, up to a few \gevc. Transverse polarization of the $\Lz$ baryon has also been observed with beams other than protons, such as $\g$, $\Kpm$, $\pipm$, $\Sigmam$, $\neum$, and neutrons on a target, as well as in semi-inclusive deep inelastic scattering~\cite{Stuntebeck:1974na, Amsterdam-CERN-Nijmegen-Oxford:1977qto, Sugahara:1979wy, K-P:1985, Gourlay:1986mf, OPAL:1997oem, NOMAD:2000wdf, WA89:2003ysd, SELEX:2007bjj, HERMES:2007fpi, HERMES:2014fmx}. For reviews of experiments and results, see for example Refs.~\cite{Bourrely:1980mr, Pondrom:1985aw, Heller:1996pg, Panagiotou:1989sv}. The observation of $\Lz$ polarization has motivated theoretical investigations over several decades to understand transverse spin effects in hadron physics ~\cite{Kane:1978nd, Panagiotou:1989sv, Dharmaratna:1996xd,Anselmino:2000vs, Boer:2009uc, Boer:2010ya, Gamberg:2021iat,Kang:2021kpt}. More recently, transverse polarization of both $\Lz$ and $\Lbar$ baryons has been observed in $\ep\en$ collisions by the \belle and BESIII collaborations~\cite{Belle:2018ttu,BESIII:2018cnd}. Belle observes that the polarization increases with the momentum fraction of the outgoing (anti)quark carried by the (anti)hyperon. Since there is no initial-state hadron in $\ep\en$ collisions, a hadronization effect must be present. 

Spontaneous polarization of different hyperons has been measured in unpolarized hadronic collisions, as reviewed in Ref.~\cite{Heller:1996pg}. A similar polarization trend as that for the $\Lz$ baryon has been observed for the $\Xiz$ and $\Xim$ states \cite{Heller:1983ia, Duryea:1991ck}, and with opposite sign for the $\Sigmap$, $\Sigmam$, and $\Sigmaz$ baryons \cite{Wilkinson:1981jy, Wah:1984swb, Dukes:1988ir}. The polarization has also been studied when the detected hyperons do not have valence quarks in common with the proton beam in unpolarized hadronic collisions, and while zero polarization was observed for the $\Lbar$ and $\Omegam$ hyperons \cite{Luk:1992ku, Heller:1978ty}, nonzero polarization was found for the $\Xibarp$ and $\Sigmabarm$ antibaryons \cite{Ho:1991rz, E761:1993qya}.

In phenomenological frameworks describing transverse $\Lz$ polarization in unpolarized collisions, the polarization originates from the nonperturbative elements in the factorization expansion of the cross-section. One well-established approach involves the polarizing transverse-momentum-dependent (TMD) fragmentation function (FF) 
\cite{Collins:2011zzd, Boer:2009uc, Boer:2010ya, Mulders:1995dh, Anselmino:2000vs}, which describes the probability that an unpolarized quark fragments into a transversely polarized hadron. In unpolarized hadronic collisions, transverse polarization can arise from the convolution of TMD parton distribution functions and TMD FFs. The polarizing TMD FF has recently been extracted through fits to the measurement by the Belle experiment \cite{Belle:2018ttu, DAlesio:2020wjq, Callos:2020qtu, Chen:2021hdn}. In hadron-hadron collisions, measurements of $\Lz$ polarization within jets offer a promising avenue to access the corresponding polarizing TMD jet FF 
\cite{Gao:2024dxl, Kang:2020xyq, Boer:2007nh,DAlesio:2024ope,Gao:2024bfp}. 

An alternative, complementary formalism is provided in the collinear approach at higher orders in the factorization expansion, which allows for the description of spin asymmetries without the need for transverse momentum or jet reconstruction. In this framework, transverse polarization arises from quark-gluon-quark and three-gluon correlators\cite{Metz:2016swz, Kanazawa:2000cx}, which encode spin-momentum correlations that can lead to a net transverse polarization of hyperons in unpolarized hadron-hadron collisions. This formalism has been used to compute contributions to the polarized hyperon cross section in unpolarized $pp$ collisions, offering insight into the mechanisms responsible for the observed $\Lz$ polarization \cite{Kanazawa:2000cx, Zhou:2008fb, Koike:2015zya, Koike:2017fxr, Ikarashi:2022zeo}. 

Hyperons decay via the weak interaction with nonconserved parity, permitting the measurement of their polarization by analyzing the angular distribution of their decay products. In the case of the $\decay{\Lz}{p\pim}$ decay, the transverse polarization is measured normal to the production plane, $\hat{n} = \hat{p}_{\text{beam}} \times \hat{p}_{\Lz}$, where $\hat{p}_{\text{beam}}$ is aligned with the beam in the direction of the LHCb detector $z$-axis \cite{Vesterinen:2014zkm}, and $\hat{p}_{\Lz}$ is the direction of the $\Lz$ momentum in the laboratory frame. The observed angular distribution for the $\Lz\to p\pi^-$ decay is given by

\begin{equation}
   \frac{1}{N} \frac{\deriv N}{\deriv\cos\theta^*} = (1+\alpha_\Lz P_{\Lz}\cos\theta^*) \etot(\cos\theta^*), 
    \label{eq:pol_fit}
\end{equation}
where $N$ is the signal yield; $\theta^*$ is the angle between $\hat{n}$ and the decay proton momentum in the $\Lz$ rest frame; $\alpha_{\Lz} = 0.746 \pm 0.009$ ($\alpha_{\Lbar} = -0.757 \pm 0.004$) \cite{PDG2024} is the parity-violating decay asymmetry; $P_{\Lz}$ is the polarization magnitude of the $\Lz$; and $\varepsilon_{\rm tot}(\cos\theta^*)$ is the total efficiency, including reconstruction, selection, and acceptance effects that are determined from a simulation without polarization effects through the detector. Polarization is determined through a binned linear fit of the efficiency-corrected $\cos\theta^*$ distribution. Similarly, for the $\Lbar$ antibaryon, the angle between $\hat{n}$ and the decay antiproton momentum is determined.

In this paper, the transverse polarization of $\Lz$ and $\Lbar$ baryons is measured as a function of \xf and \pt in $p$Pb collisions at the center-of-mass energy in the nucleon-nucleon system $\sqsnn=5.02\tev$. The $p$Pb measurement covers the backward rapidity range $-5.0 < y^* < -2.5 $, in which the lead beam enters the LHCb detector at the interaction point, and the forward rapidity range $1.5 < y^* < 4.0$, where the proton beam enters the LHCb detector at the interaction point. Here, $y^*$ is defined in the nucleon-nucleon center-of-mass system and is related to the rapidity in the laboratory frame by $y^* = y_{\text{lab}} - 0.465$ in the forward configuration, and $y^* = -(y_{\text{lab}} + 0.465)$ in the backward configuration. This measurement covers a poorly explored kinematic range and a much higher center-of-mass energy per nucleon than previous results, probing the $\Lz$ and $\Lbar$ transverse polarization for $0.15 < \pt < 6.00 \gevc$, as well as $0.005 < \xf < 0.040$ in the forward configuration and $-0.10 < \xf < -0.01$ in the backward configuration. In symmetric collisions, such as $pp$, the polarization must be zero at $\xf = 0$ and an antisymmetric function of \xf, $P(-\xf) = -P(\xf)$. In measurements using $p$Pb collisions, the polarization is expected to be consistent with zero at $\xf\approx0$, as observed in previous experiments ~\cite{ATLAS:2014ona, HERA-B:2006rds, LHCb-PAPER-2024-009}, and any nonzero polarization is likely to arise from nuclear effects.

\section{Detector, data, and simulation}
The LHCb detector~\cite{LHCb-DP-2008-001,LHCb-DP-2014-002} is a single-arm forward spectrometer covering the pseudorapidity range $2 < \eta < 5$. The detector elements that are particularly relevant to this analysis are a silicon-strip vertex detector surrounding the $p$Pb interaction region; a tracking system that provides a measurement of the momentum, $p$, of charged particles; and two ring-imaging Cherenkov (RICH) detectors that can discriminate between different species of charged hadrons. The polarity of the detector dipole magnet is reversed periodically during data taking, and the corresponding datasets are added together for this measurement. The corresponding integrated luminosity for the forward (backward) $p$Pb data sample used for these results is $1.14 \pm 0.02 \invnb$ ($ 0.45 \pm 0.01 \invnb)$. Events are selected by a minimum-bias trigger that requires at least one reconstructed track in the vertex detector. A further requirement of at least one reconstructed primary vertex (PV) is imposed. Simulated events are used to model the effects of the detector acceptance and the selection requirements. In the simulation, $p$Pb collisions are generated using EPOS-LHC event generator with a specific \lhcb configuration~\cite{Pierog:2013ria}, with $15\times10^6$ events generated for both the forward and backward rapidity. Decays of unstable particles are described by \evtgen~\cite{Lange:2001uf}, and the interaction of the generated particles with the detector, and its response, are implemented using the \geant toolkit~\cite{Agostinelli:2002hh,Allison:2006ve} as described in Ref.~\cite{LHCb-PROC-2011-006}. A simulation sample is used for each collision configuration.

\section{Polarization measurement}
\subsection{Signal selection and yields extraction}
To reconstruct the $\Lz$ and $\Lbar$ candidates, pairs of oppositely charged proton and pion candidates are selected and required to traverse all tracking detectors, have a good quality track fit, momentum greater than $5\gevc$, and a common good-quality vertex. The particles must have a combined invariant mass $1100 < m(\proton\pim) < 1135 \mevcc$. 
Candidates are required to have a longitudinal position of the associated PV in the range \mbox{$-180 < z_{\rm PV}
< 180$\mm}. A particle identification (PID) requirement based on a neural network trained to separate particle types using information from several detectors, including the RICH system, is applied to select protons. A discriminant, $\nu$, is used to increase the signal-to-background ratio, following previous \Lz production studies in LHCb~\cite{LHCb-PAPER-2011-005}. The discriminant is based on the logarithm of the $\chisqip$, defined as the difference in the vertex-fit $\chisq$ of a given PV reconstructed with and without the particle under consideration. The variable is defined as $\nu = \log(I^+I^-/I^0)$, where $I^\pm$ is the $\chisqip$ of the proton or pion particle, and $I^0$ is the $\chisqip$ of the $\Lz$ or $\Lbar$ candidate. This requirement selects prompt contributions of $\Lz$ or $\Lbar$ candidates produced directly in the collision as well as those from short-lived ($c\tau < 1\nm$) decays. With this selection and definition, the signal candidates are 98\% prompt, which includes a 22\% feed-down contribution from $\Sigmaz\to\Lz\gamma$ decays estimated from the EPOS-LHC simulation. According to the SU(6) quark model, the \Sigmaz and \Lz baryons are produced with opposite polarization orientations and differing polarization magnitudes following the relation $P_\Sigmaz = -\frac{1}{3}P_\Lz$ \cite{Andersson:1979wj,Andersson:1983ia,Gustafson:1992iq}, a relation that had also been predicted earlier in Ref.~\cite{PhysRev.140.B668}. If the polarization of prompt \Lz and \Sigmaz baryons originate from the strange-quark spin, the feed-down contributions from \Sigmaz\to\Lz\g decays is expected to reduce the magnitude of the observed \Lz polarization.

\begin{figure*}\centering\includegraphics[width=0.49\linewidth]{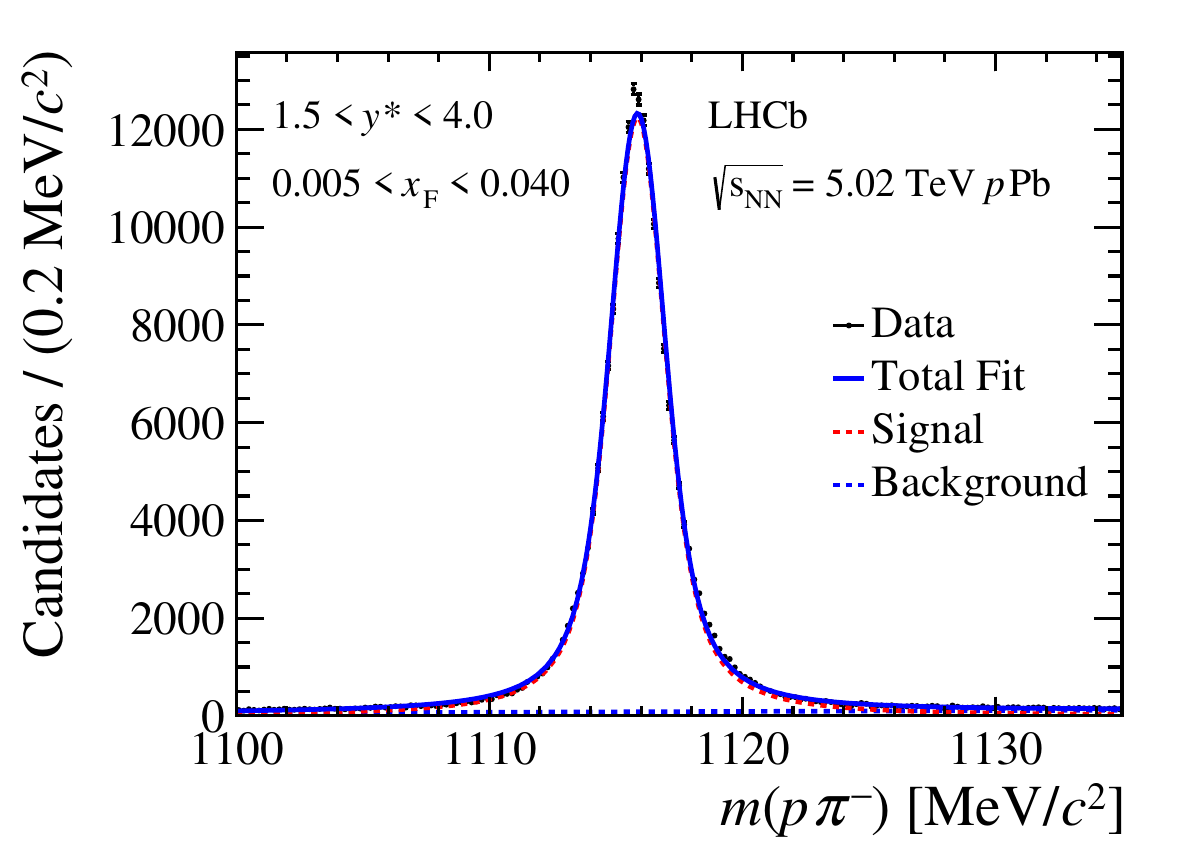}
    \includegraphics[width=0.49\linewidth]{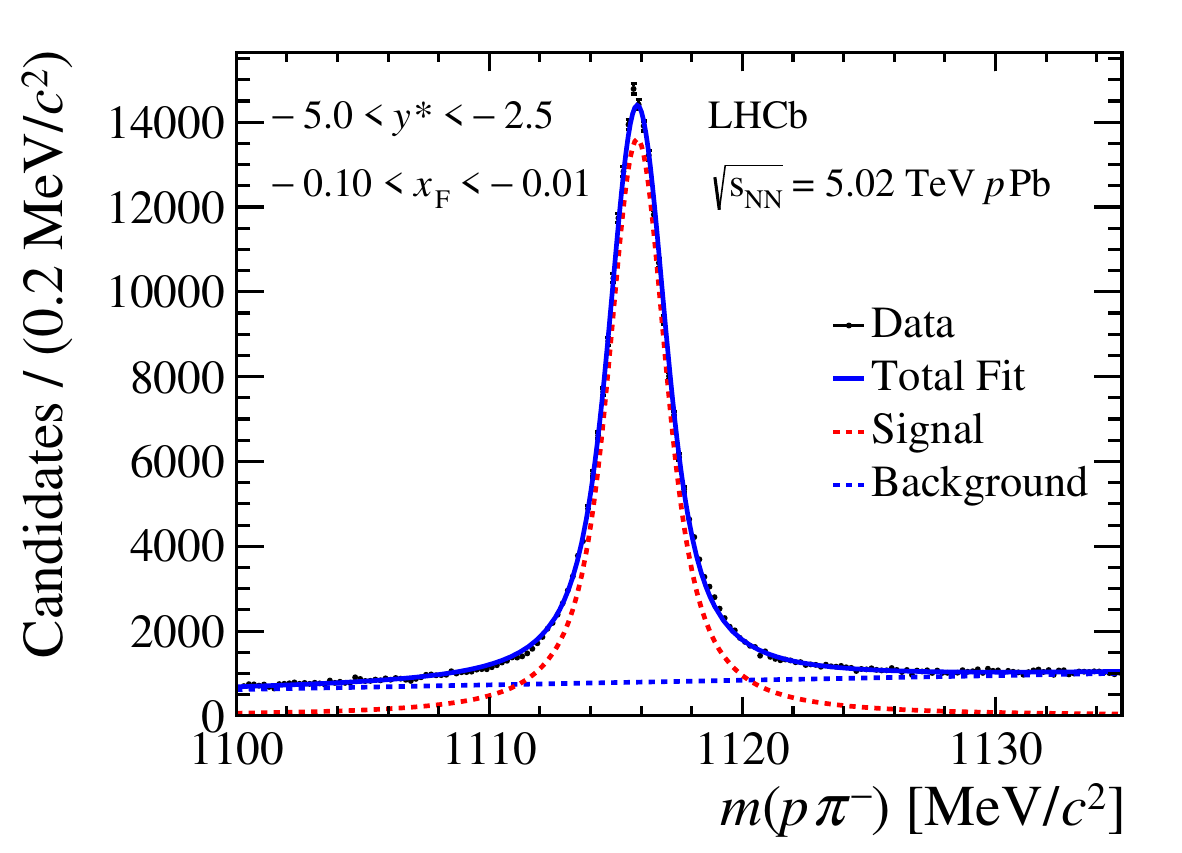}
        \includegraphics[width=0.49\linewidth]{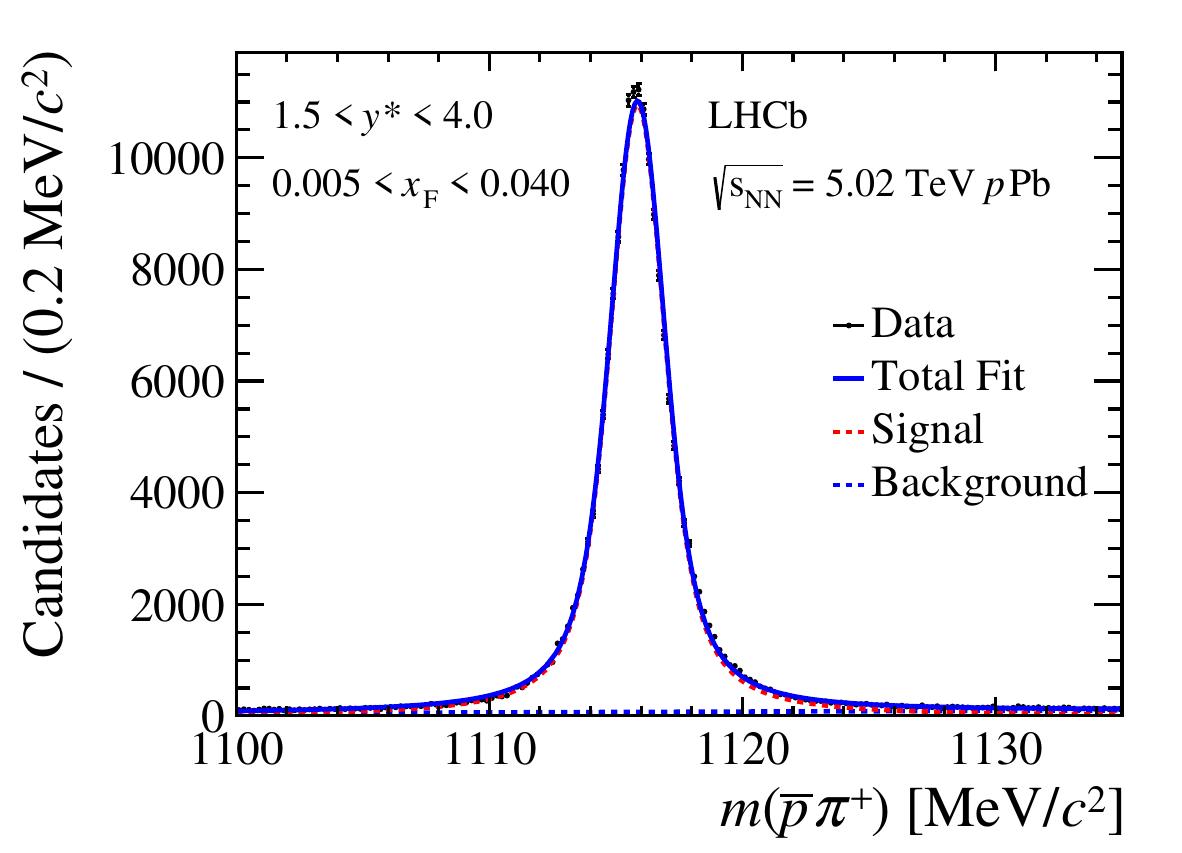}
            \includegraphics[width=0.49\linewidth]{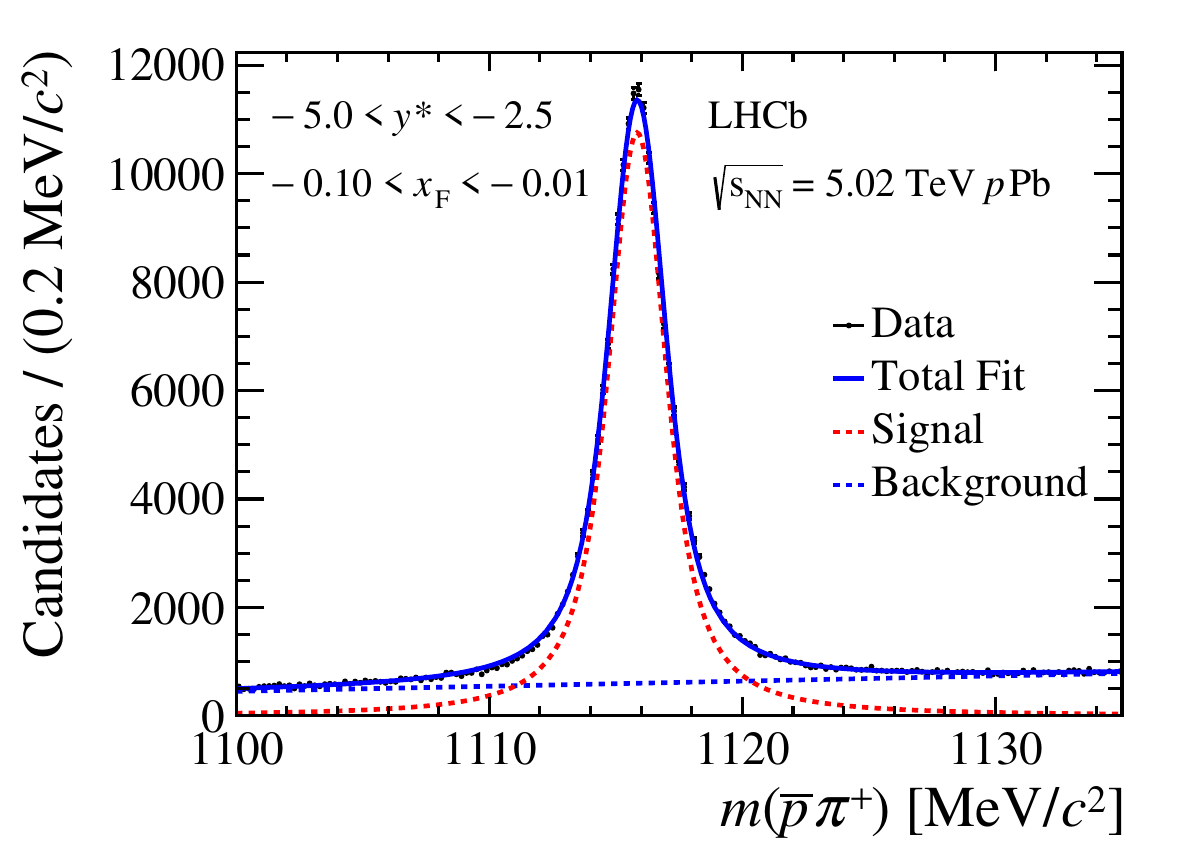}
    \caption{Invariant-mass distribution for $\Lz$ (top) and $\Lbar$ (bottom) candidates and fit results in data after all selection requirements for $0.15 < \pt < 6.00 \gevc$ in two intervals of $y^*$ and \xf.}
    \label{fig:fittedmass}
\end{figure*}

The invariant-mass distributions for the selected $\Lz$ and $\Lbar$ candidates are shown in Fig.~\ref{fig:fittedmass}. There are 231\,358 (205\,590) \Lz (\Lbar) and 396\,045 (307\,853) \Lz (\Lbar) candidates in the signal region, $1111 < m_{\proton\pi} < 1121 \mevcc$, at positive and negative $y^*$, respectively. The distributions are fitted using a binned, extended maximum-likelihood fit, where the signal component is modeled by a Voigtian function, a convolution of a Breit--Wigner with a Gaussian function. The mean, Breit--Wigner width, and Gaussian resolution are treated as free parameters, initialized to values estimated from simulation and constrained within bounds to ensure fit stability. Similarly, the background is modeled using a first-order Bernstein polynomial with two free parameters. Altogether, the fit includes seven free parameters: three for signal shape, two for background shape, and two for signal and background yields, all optimized simultaneously during the fit. The signal purity is about 97\% and 85\% in the signal region at positive and negative $y^*$, respectively, with the higher background at negative $y^*$ due to the higher-multiplicity events when the lead beam enters the LHCb detector at the interaction point.

\subsection{Efficiency determination}
\begin{figure*}
    \centering
    \includegraphics[width=\linewidth]{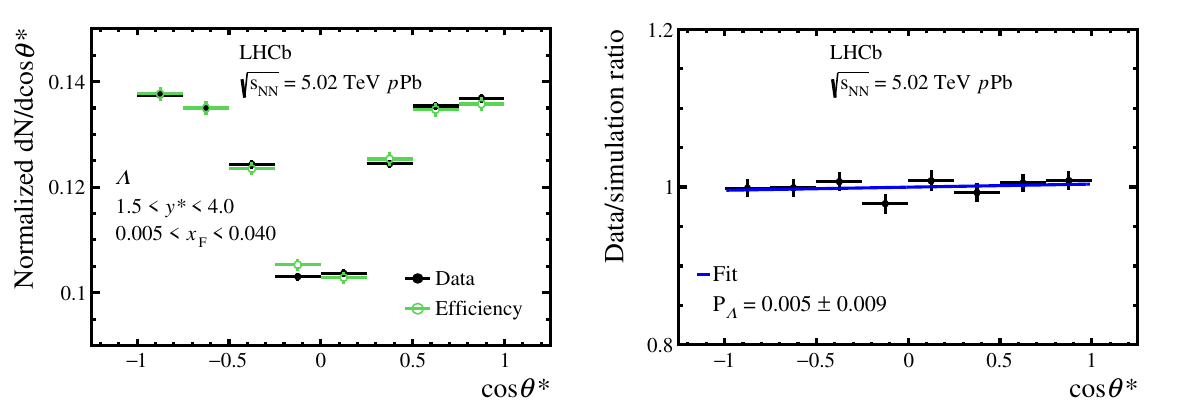}
\caption{Left: decay angle distribution measured in data and the total efficiency for $\Lz$ production in $1.5 < y^* < 4.0$ and $0.15 < \pt < 6.00 \gevc$, where both have a common normalization. Right: corrected-data $\cos\theta^*$ distribution, along with the linear fit to extract the polarization. The fit reduced $\chisq$ is $4.2/6$ and the error bars represent the sum in quadrature of the statistical uncertainties of data and simulation.}
    \label{fig:costheta_forward}
\end{figure*}

The total efficiency $\etot$ in Eq.~(\ref{eq:pol_fit}) includes the effects of geometric acceptance, the reconstruction and selection, and the PID criteria. The (anti)baryons generated in the EPOS-LHC simulated sample are unpolarized and, therefore, correspond to a uniform $\cos\theta^*$ distribution, which is affected by the detector acceptance and selection criteria. The signal yields in the simulated sample are derived from reconstructed candidates that are matched to generated (truth-level) particles following a standard matching algorithm. The simulation is adjusted using weights to improve agreement with data. The weight is calculated as the ratio of the normalized distributions of background-subtracted data and simulated signal in two-dimensional bins of \pt and $\eta$ for the $\Lz$ and $\Lbar$ baryons, and the independent variable $z_{\rm PV}$. The weighting procedure has no impact on the $\cos\theta^*$ distribution.
The PID efficiency is estimated in both simulation and data and computed as a function of \pt and $\eta$ of the $\Lz$ and $\Lbar$. The ratio of the efficiencies is applied as a weight to the simulation. The track-reconstruction efficiency from the simulation is corrected with a tag-and-probe method applied to data in two-dimensional intervals of $\eta$ and \pt using a calibration sample of $\jpsi\to\mup\mun$ decays in the range $5 < p < 200 \gevc$ \cite{LHCb-DP-2013-002}. Figure~\ref{fig:costheta_forward} (left) shows the decay angle distribution for protons from $\Lz$ baryons measured in data, along with the respective total efficiencies for $1.5 < y^* < 4.0$ and $0.15 < \pt < 6.00 \gevc$. Each point is obtained by extracting the signal yield in the given bin using an independent fit to the invariant-mass distribution. The efficiency-corrected angular distribution and the fit to a linear function, $p_0(1+p_1\cos\theta^*)$, where $p_0$ and $p_1$ are free parameters, and $P_{\Lz} = p_1/\alpha_{\Lz}$, are shown on the right.

\subsection{Systematic uncertainties}
Several sources of systematic uncertainty are considered, and evaluated as the difference in the extracted polarization between the baseline and various alternative approaches. The systematic uncertainties related to signal modeling are estimated by performing an alternative fit represented by a Voigtian combined with an additional Gaussian distribution which introduces another resolution term as a free parameter in the fit. The uncertainty from the background modeling is evaluated by replacing the Bernstein polynomial with a first-degree polynomial with two free parameters in the fit. The uncertainty from the calculation of the simulation weights is estimated by calculating the weights using all particle candidates after selection requirements, without subtracting the background. Additionally, the effect of detector occupancy is taken into account by applying a weight to the simulation according to the track multiplicity. The uncertainty associated with the estimation of PID efficiencies is evaluated using an alternative method, in which the PID efficiency is calculated as the single-track proton efficiency in bins of proton momentum and $\eta$. Since the angular efficiency is derived from truth-matched yields and is sensitive to reduced matching performance for lower-quality tracks, a correction is applied to account for reconstructed candidates in simulation that fail to match a generated particle. The truth-matching efficiency is defined as the ratio of matched to total signal yields in simulation, where the latter also includes a fitted unmatched component determined from the invariant-mass distribution. 

The correction is applied in two-dimensional bins of the angular distribution with \pt and \xf, and the associated systematic uncertainty is taken as the difference in the polarization result with and without this correction. The impact of the binning choice for the angular distribution used in the fit is also considered by repeating the polarization measurement using five bins instead of eight. The impact of the selection of $\nu$ is studied by varying the selection criteria above and below the baseline value. An additional uncertainty on the efficiency determination arises from the size of the simulation sample and is calculated by varying the simulated distribution by its statistical uncertainty. Other sources of uncertainty, such as the uncertainty on $\alpha_{\Lz}$ or $\alpha_{\Lbar}$ and the precession of the $\Lz$ or $\Lbar$ spin as it traverses the magnetic field, are also considered but found to be negligible. The track reconstruction efficiency has an uncertainty due to the size of the calibration samples and the modeling of hadron material interactions, and is also found to be negligible. The total bin-dependent systematic uncertainty is computed in quadrature in bins of \xf and \pt. The systematic uncertainties for the integrated polarization values are shown in Table \ref{tab:uncertainty}.

Cross-checks are performed to investigate potential biases in the measurement. The polarization of the \KS meson, a pseudoscalar with null polarization, is measured using the same method and found to be consistent with zero ($0.000 \pm 0.003$ and $0.000 \pm 0.002$ for the forward ($1.5 < y^* < 4.0$) and backward ($-5.0 < y^* < -2.5$) collision configurations, where the uncertainties are statistical only). The polarization is additionally calculated separately for each magnet polarity before combining, and the results are found to be consistent. A potential nonlinearity of the angular distribution is also considered by adding a quadratic term to the fitting model, which would be sensitive to residual imperfections in the modeling of the efficiency. No significant bias is observed in these cross-checks.

\begin{table}[]
    \centering
    \caption{Systematic uncertainty contributions for the forward ($1.5 < y^* < 4.0$) and backward ($-5.0 < y^* < -2.5$) collision configurations. The numbers represent the absolute systematic uncertainties of the integrated polarization values. }
    
\begin{tabular}{ccccc}
     \multicolumn{1}{c}{}& \multicolumn{4}{c}{Systematic uncertainty}                                      \\ \hline

     \multicolumn{1}{c}{}& \multicolumn{2}{c}{Forward}  & \multicolumn{2}{c}{Backward}                                      \\ 
\multicolumn{1}{l}{Source} & \multicolumn{1}{c}{$\Lz$} & \multicolumn{1}{c}{$\Lbar$} & \multicolumn{1}{c}{$\Lz$} & \multicolumn{1}{c}{$\Lbar$} \\ \hline

\multicolumn{1}{l}{Signal model} & \multicolumn{1}{c}{0.000} & \multicolumn{1}{c}{0.000} & \multicolumn{1}{c}{0.001} & \multicolumn{1}{c}{0.000} \\

\multicolumn{1}{l}{Background model} & \multicolumn{1}{c}{0.003} & \multicolumn{1}{c}{0.001} & \multicolumn{1}{c}{0.003} & \multicolumn{1}{c}{0.001} \\

\multicolumn{1}{l}{Kinematic weights} & \multicolumn{1}{c}{0.000} & \multicolumn{1}{c}{0.000} & \multicolumn{1}{c}{0.005} & \multicolumn{1}{c}{0.002} \\

\multicolumn{1}{l}{Simulation size} & \multicolumn{1}{c}{0.005} & \multicolumn{1}{c}{0.005} & \multicolumn{1}{c}{0.004} & \multicolumn{1}{c}{0.005} \\

\multicolumn{1}{l}{PID efficiencies} & \multicolumn{1}{c}{0.002} & \multicolumn{1}{c}{0.002} & \multicolumn{1}{c}{0.005} & \multicolumn{1}{c}{0.001} \\

\multicolumn{1}{l}{Truth matching efficiency} & \multicolumn{1}{c}{0.002} & \multicolumn{1}{c}{0.002} & \multicolumn{1}{c}{0.001} & \multicolumn{1}{c}{0.001} \\

\multicolumn{1}{l}{Binning of $\cos\theta^*$ distribution} & \multicolumn{1}{c}{0.003} & \multicolumn{1}{c}{0.002} & \multicolumn{1}{c}{0.003} & \multicolumn{1}{c}{0.000} \\

\multicolumn{1}{l}{Selection requirement on $\nu$} & \multicolumn{1}{c}{0.002} & \multicolumn{1}{c}{0.002} & \multicolumn{1}{c}{0.001} & \multicolumn{1}{c}{0.002} \\

\multicolumn{1}{l}{Multiplicity dependence} & \multicolumn{1}{c}{0.000} & \multicolumn{1}{c}{0.000} & \multicolumn{1}{c}{0.000} & \multicolumn{1}{c}{0.001} \\ \hline

\multicolumn{1}{l}{Total} & \multicolumn{1}{c}{0.007} & \multicolumn{1}{c}{0.006} & \multicolumn{1}{c}{0.009} & \multicolumn{1}{c}{0.006} \\ \hline

\end{tabular}
    \label{tab:uncertainty}
\end{table}

\subsection{Results}
The transverse polarization is determined by fitting the efficiency-corrected decay angle distribution based on Eq.~(\ref{eq:pol_fit}). The polarization is measured for $0.15 < \pt < 6.00 \gevc$, as well as for $0.005 < \xf < 0.040$ in the forward ($1.5 < y^* < 4.0$) configuration and $-0.10 < \xf < -0.01$ in the backward ($-5.0 < y^* < -2.5$) configuration. Figure~\ref{fig:costheta_forward} (right) shows the $\cos\theta^*$ distribution after acceptance corrections integrated over the entire kinematic range for $\Lz$ baryons in the forward configuration. The linear fit and the corresponding extracted polarization value are shown, where the error bars represent the sum in quadrature of the data and simulation statistical uncertainties. The integrated polarization measurements for $1.5 < y^* < 4.0$ are
\begin{align*}
    P_{\Lz} &= 0.005 \pm 0.008 \pm 0.007, \\
     P_{\Lbar} &= 0.007 \pm 0.009 \pm 0.006. 
\end{align*} 
The integrated polarization measurements for $-5.0 < y^* < -2.5$ are
\begin{align*}
    P_{\Lz} &= 0.004 \pm 0.008 \pm 0.009, \\
     P_{\Lbar} &= -0.005 \pm 0.008 \pm 0.006. 
\end{align*} 

\begin{figure*}
    \centering
    \includegraphics[width=0.49\linewidth]{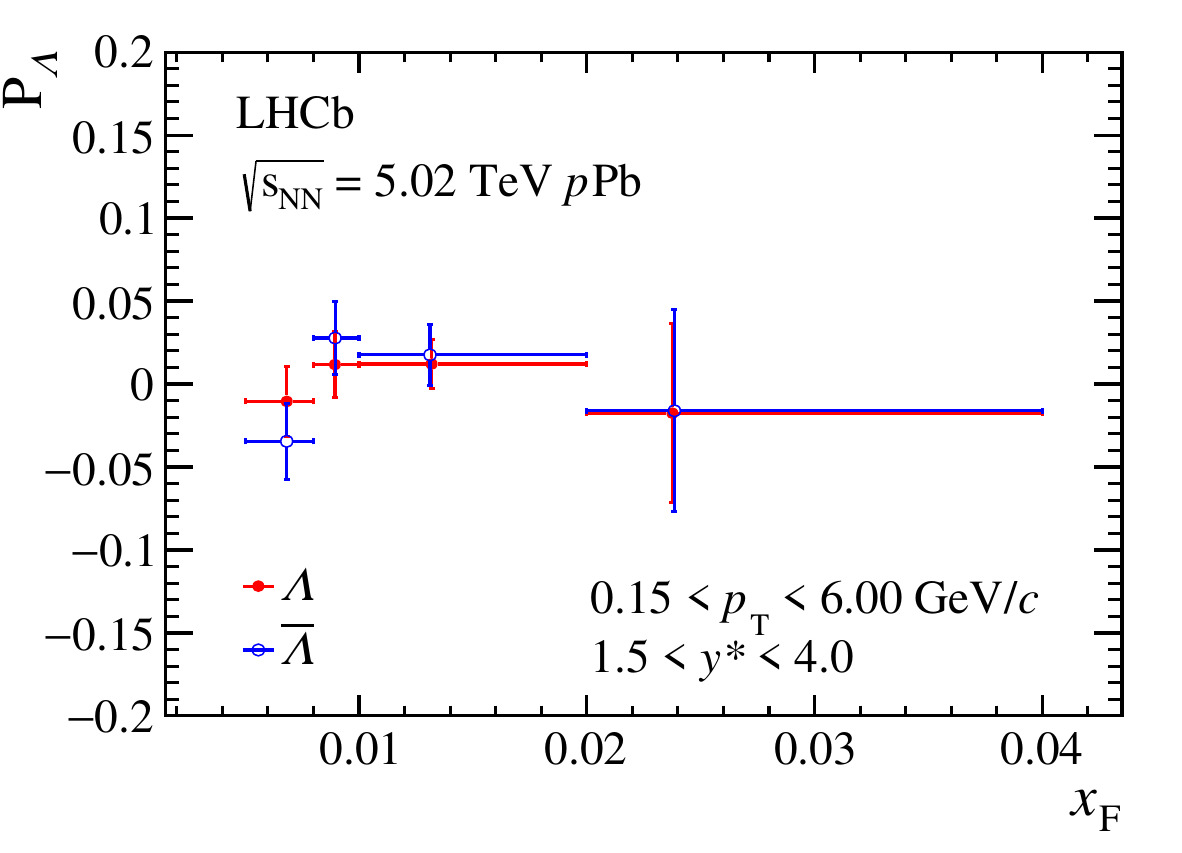}
    \includegraphics[width=0.49\linewidth]{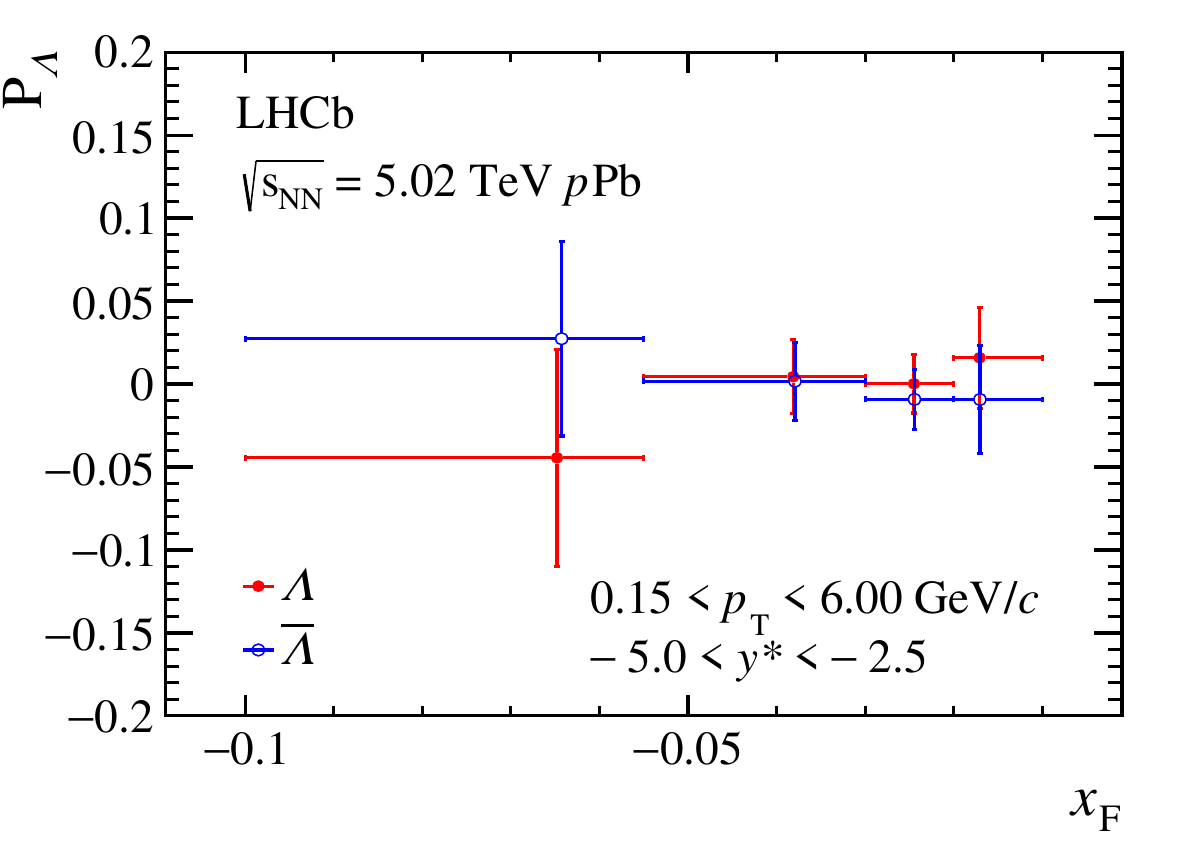}
    \caption{Transverse polarization for $\Lz$ (red filled markers) and $\Lbar$ (blue empty markers) baryons as a function of Feynman-$x$ for (left) forward ($1.5 < y^* < 4.0$) and (right) backward ($-5.0 < y^* < -2.5$) collision configurations. The error bars represent the total uncertainty. }
    \label{fig:pol_vs_xF}
\end{figure*}

\begin{figure*}
    \centering
    \includegraphics[width=0.49\linewidth]{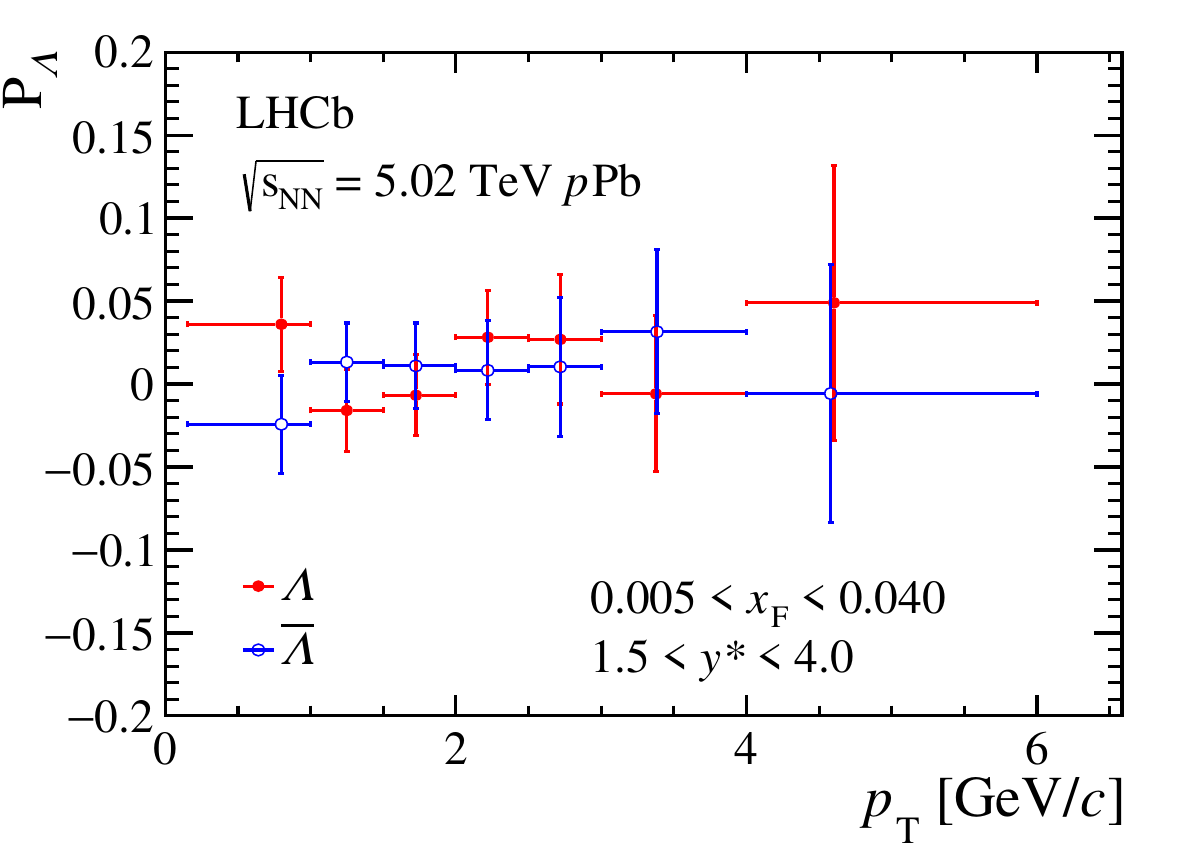}
    \includegraphics[width=0.49\linewidth]{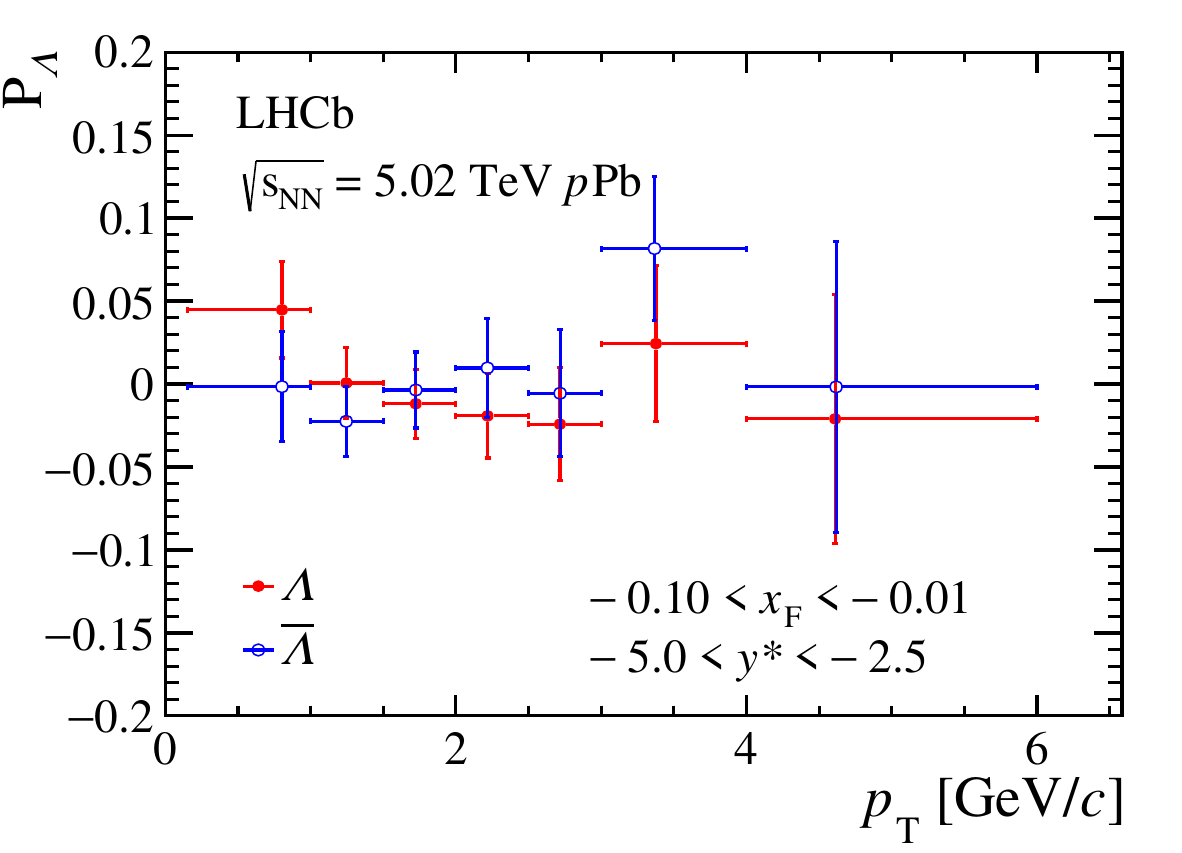}
    \caption{Transverse polarization for $\Lz$ (red, filled) and $\Lbar$ (blue, empty) as a function of transverse momentum for (left) forward and (right) backward collision configurations. The error bars represent the total uncertainty. }
    \label{fig:pol_vs_pT}
\end{figure*}

The polarization is also determined in bins of \pt and \xf. The results are summarized in Tables \ref{table:LzpPbPol}, \ref{table:LbarpPbPol}, \ref{table:LzPbpPol}, and \ref{table:LbarPbpPol}, where the first uncertainties are statistical from data, and the second represent the systematic uncertainties added in quadrature. Figures~\ref{fig:pol_vs_xF} and \ref{fig:pol_vs_pT} show the results as a function of \xf and \pt, respectively. The statistical and systematic uncertainties are added in quadrature to obtain the total uncertainty. All measured polarization values are consistent with zero.

\begin{figure*}
    \centering
    \includegraphics[width=0.49\linewidth]{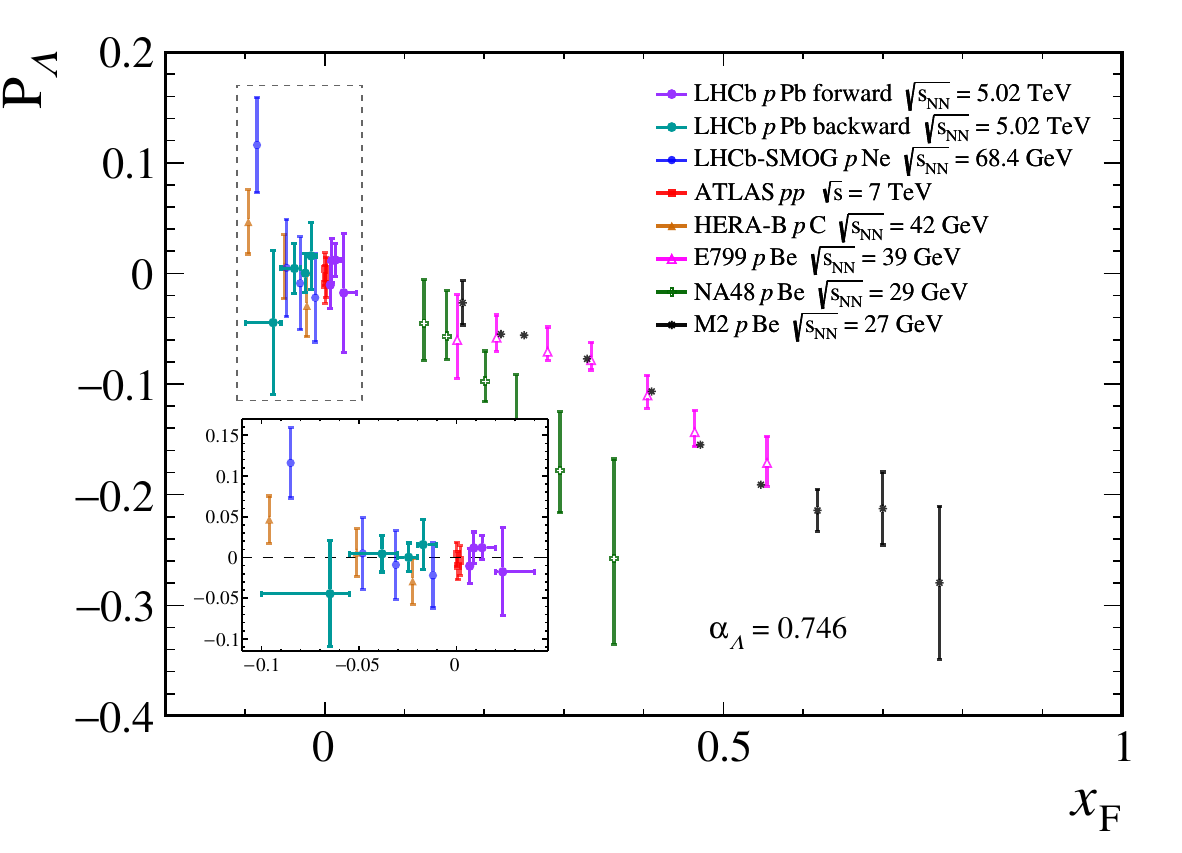}
    \includegraphics[width=0.49\linewidth]{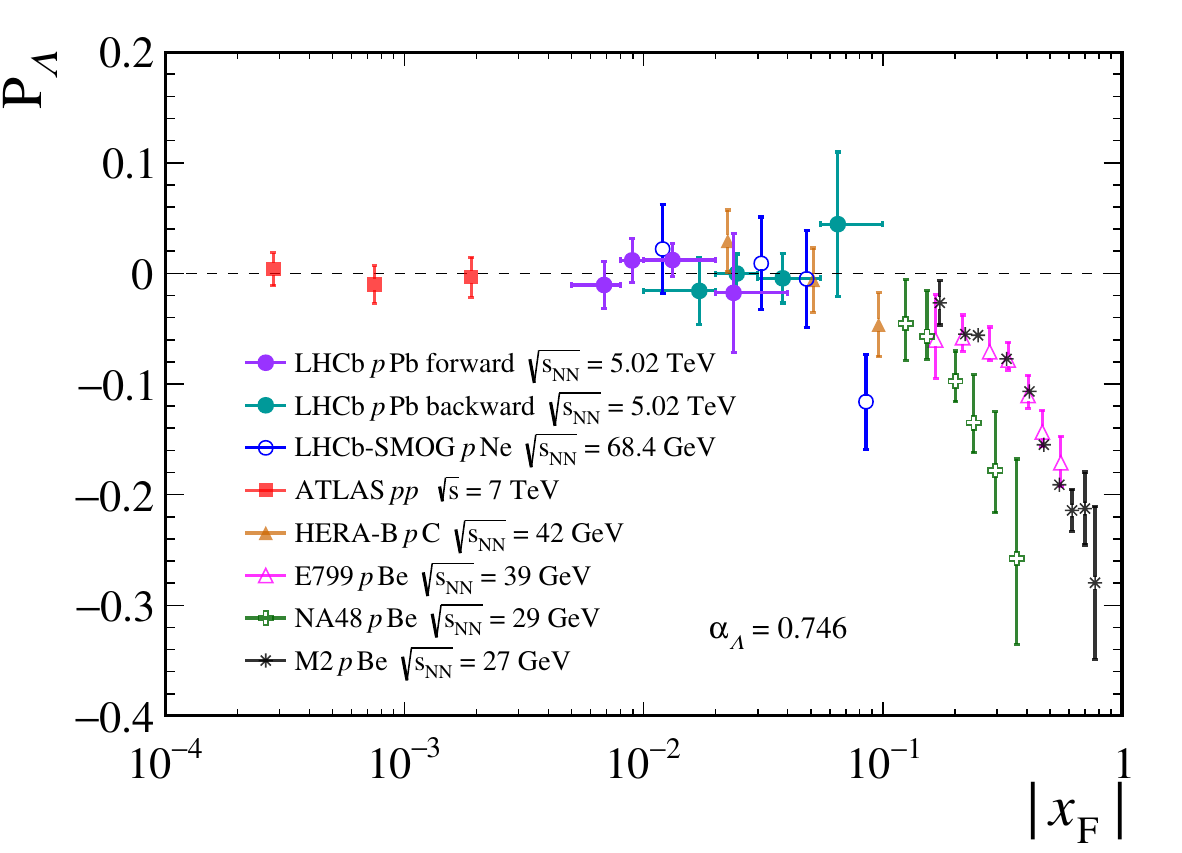}
    \caption{ Left: transverse $\Lz$ polarization measured in LHCb $p$Pb collisions in the forward \mbox{($1.5 < y^*< 4.0$)} and backward ($-5.0 < y^*< -2.5$) configurations and LHCb-SMOG fixed-target $p$Ne collisions~\cite{LHCb-PAPER-2024-009}, compared with other measurements with different collision systems and different center-of-mass energies~\cite{Lundberg:1989hw,Ramberg:1994tk,Fanti:1998px,HERA-B:2006rds,ATLAS:2014ona}. The values have been scaled to match the $\alpha_{\Lz}$ value used in this paper.  Right: transverse $\Lz$ polarization comparison with previous measurements, assuming the transformation $P(-\xf) = -P(\xf)$. }
    \label{fig:comparison}
\end{figure*}

Figure~\ref{fig:comparison} (left) shows the transverse $\Lz$ polarization measured in $p$Pb collisions at $\sqsnn=5.02\tev$ for positive and negative \xf, together with earlier results from fixed-target $p$Ne collision measured by LHCb-System for Measuring the Overlap with Gas (SMOG)~\cite{LHCb-PAPER-2024-009} and from other unpolarized $pp$ and $p$A experiments, including the M2 beamline~\cite{Lundberg:1989hw} and the E799 experiment~\cite{Ramberg:1994tk} at Fermilab, NA48~\cite{Fanti:1998px} and ATLAS experiment at the CERN Large Hadron Collider~\cite{ATLAS:2014ona}, and HERA-B experiment~\cite{HERA-B:2006rds} at Deutsches Elektronen-Synchrotron (DESY). The results have been rescaled to the $\alpha_\Lz$ value used in this analysis and, where necessary, the \xf values were converted from the laboratory frame to the center-of-mass. The compared data span a wide range of collision energies, from fixed-target experiments at 27\gev up to LHC measurements of 68\gev, 5 and 7\tev, and include different collision systems ($pp$, $p$Be, $p$C, $p$Ne, and $p$Pb). The fixed-target experiments cover lower \pt ranges ($0.3$--$2\gevc$), while ATLAS and LHCb extend to higher \pt. The fixed-target results show negative polarization, reaching $P_\Lz \approx -0.3$ for $\xf \gtrsim 0.2$ and decreasing in magnitude as \xf approaches zero. A comparison of E799 ($0.67 < \pt < 2.15 \gevc$) and NA48 ($0.28 < \pt < 0.86 \gevc$) at overlapping \xf indicates a possible \pt dependence of the polarization magnitude. The present $p$Pb results show polarization consistent with zero ($|P_\Lz| \leq 0.05$) over the measured $\xf$ range and $\pt$ interval of $0.15$--$6.00\gevc$. Figure~\ref{fig:comparison} (right) shows on a logarithmic $\xf$ scale the same comparison with the backward-rapidity results transformed to positive \xf values using the relation $P(-\xf) = -P(\xf)$, which must hold for symmetric collision systems such as $pp$, to include also previous results. In the asymmetric $p$Pb collisions, possible deviations from zero at low $\xf$ could arise from differences in interactions in the target or other nuclear effects; however, the present results show no significant deviation from zero within the uncertainties. The present LHCb measurements reach an average \pt up to 4.6\gevc, entering a perturbative regime where additional theoretical tools can be applied.

\begin{table}[tb]
    \centering
    \caption{Polarization measured in the integrated kinematic range and in \xf and \pt bins for \Lz in the forward ($1.5 < y^* < 4.0$) configuration. }
\label{table:LzpPbPol}
\begin{tabular}{lccr}
 Forward \Lz & $\langle\xf\rangle$ & $\langle\pt\rangle$  & \multicolumn{1}{c}{Polarization} \\ 
& $(10 ^{-2})$ & (\gevc) &\\ \hline
Integrated & 1.1	&	1.6	&	0.005	$\pm$	0.008	$\pm$	0.007 \\ \hline
\xf ($\times 10^{-2}$)   & & &   \\ 
$[0.5 , 0.8]$	&	0.7	&	1.3	&	$-0.010$ $\pm$ 0.016 $\pm$ 0.013 \\
$[0.8 , 1.0]$	&	0.9	&	1.5	&	0.012 $\pm$ 0.016 $\pm$ 0.011 \\
$[1.0 , 2.0]$	&	1.3	&	1.8	&	0.012 $\pm$ 0.012 $\pm$ 0.008 \\
$[2.0 , 4.0]$	&	2.4	&	2.8	&	$-0.018$ $\pm$ 0.040 $\pm$ 0.036 \\

 \hline
\pt (\gevc)   & & &  \\ 
$[0.15 , 1.00]$	&	0.8	&	0.8	&	0.036 $\pm$ 0.019 $\pm$ 0.021 \\
$[1.00 , 1.50]$	&	0.9	&	1.2	&	$-0.016$ $\pm$ 0.015 $\pm$ 0.020 \\
$[1.50 , 2.00]$	&	1.1	&	1.7	&	$-0.007$ $\pm$ 0.017 $\pm$ 0.018 \\
$[2.00 , 2.50]$	&	1.3	&	2.2	&	0.028 $\pm$ 0.022 $\pm$ 0.017 \\
$[2.50 , 3.00]$	&	1.4	&	2.7	&	0.027 $\pm$ 0.034 $\pm$ 0.020 \\
$[3.00 , 4.00]$	&	1.5	&	3.4	&	$-0.006$ $\pm$ 0.036 $\pm$ 0.030 \\
$[4.00 , 6.00]$	&	1.6	&	4.6	&	0.049 $\pm$ 0.066 $\pm$ 0.049 \\
\hline
    \end{tabular}

\end{table}

\begin{table}[tb]
    \centering
\caption{Polarization measured in the integrated kinematic range and in \xf and \pt bins for \Lbar in the forward ($1.5 < y^* < 4.0$) configuration. }
\label{table:LbarpPbPol}
\begin{tabular}{lccr}
  Forward \Lbar & $\langle\xf\rangle$ & $\langle\pt\rangle$  & \multicolumn{1}{c}{Polarization} \\ 
& $(10 ^{-2})$ & (\gevc) &\\ \hline
Integrated & 1.0	&	1.6	&	0.007	$\pm$	0.009 $\pm$	0.006
\\ \hline
\xf ($\times 10^{-2}$)   & & &   \\ 
$[0.5 , 0.8]$	&	0.7	&	1.3	&	$-0.035$ $\pm$ 0.019 $\pm$ 0.014 \\
$[0.8 , 1.0]$	&	0.9	&	1.5	&	0.028 $\pm$ 0.019 $\pm$ 0.012 \\
$[1.0 , 2.0]$	&	1.3	&	1.8	&	0.018 $\pm$ 0.013 $\pm$ 0.013 \\
$[2.0 , 4.0]$	&	2.4	&	2.8	&	$-0.016$ $\pm$ 0.049 $\pm$ 0.038 \\

 \hline
\pt (\gevc)   & & &  \\ 
$[0.15 , 1.00]$	&	0.8	&	0.8	&	$-0.024$ $\pm$ 0.021 $\pm$ 0.020 \\
$[1.00 , 1.50]$	&	0.9	&	1.2	&	0.013 $\pm$ 0.015 $\pm$ 0.018 \\
$[1.50 , 2.00]$	&	1.1	&	1.7	&	0.011 $\pm$ 0.017 $\pm$ 0.019 \\
$[2.00 , 2.50]$	&	1.2	&	2.2	&	0.008 $\pm$ 0.024 $\pm$ 0.017 \\
$[2.50 , 3.00]$	&	1.4	&	2.7	&	0.010 $\pm$ 0.034 $\pm$ 0.025 \\
$[3.00 , 4.00]$	&	1.5	&	3.4	&	0.031 $\pm$ 0.040 $\pm$ 0.030 \\
$[4.00 , 6.00]$	&	1.6	&	4.6	&	$-0.006$ $\pm$ 0.068 $\pm$ 0.039 \\
\hline
    \end{tabular}
\end{table}

\begin{table}[tb]
    \centering
\caption{Polarization measured in the integrated kinematic range and in \xf and \pt bins for \Lz in the backward ($-5.0 < y^* < -2.5$) configuration. }
\label{table:LzPbpPol}
    
\begin{tabular}{lccr}
 Backward \Lz & $\langle\xf\rangle$ & $\langle\pt\rangle$  & \multicolumn{1}{c}{Polarization} \\ 
& $(10 ^{-2})$ & (\gevc) &\\ \hline
Integrated & $-2.7$	&	1.6	&	0.004	$\pm$	0.008 $\pm$	0.009  \\ \hline
\xf ($\times 10^{-2}$)   & & &   \\ 
$[-10.0 , -5.5]$	&	$-6.5$	&	2.8	&	$-0.044$ $\pm$ 0.042 $\pm$ 0.050 \\
$[-5.5 , -3.0]$	&	$-3.8$	&	1.9	&	0.004 $\pm$ 0.013 $\pm$ 0.018 \\
$[-3.0 , -2.0]$	&	$-2.4$	&	1.5	&	0.000 $\pm$ 0.012 $\pm$ 0.013 \\
$[-2.0 , -1.0]$	&	$-1.7$	&	1.3	&	0.016 $\pm$ 0.018 $\pm$ 0.025 \\

\hline
\pt (\gevc)   & & &  \\ 
$[0.15 , 1.00]$	&	$-2.1$	&	0.8	&	0.045 $\pm$ 0.021 $\pm$ 0.020 \\
$[1.00 , 1.50]$	&	$-2.5$	&	1.2	&	0.001 $\pm$ 0.013 $\pm$ 0.016 \\
$[1.50 , 2.00]$	&	$-2.9$	&	1.7	&	$-0.012$ $\pm$ 0.015 $\pm$ 0.015 \\
$[2.00 , 2.50]$	&	$-3.4$	&	2.2	&	$-0.019$ $\pm$ 0.020 $\pm$ 0.015 \\
$[2.50 , 3.00]$	&	$-3.7$	&	2.7	&	$-0.024$ $\pm$ 0.028 $\pm$ 0.019 \\
$[3.00 , 4.00]$	&	$-4.0$	&	3.4	&	0.024 $\pm$ 0.034 $\pm$ 0.033 \\
$[4.00 , 6.00]$	&	$-4.3$	&	4.6	&	$-0.021$ $\pm$ 0.063 $\pm$ 0.040 \\
\hline
    \end{tabular}
\end{table}

\begin{table}[tb]
    \centering
\caption{Polarization measured in the integrated kinematic range and in \xf and \pt bins for \Lbar in the backward ($-5.0 < y^* < -2.5$) configuration. }
\label{table:LbarPbpPol}
\begin{tabular}{lccr}
Backward \Lbar & $\langle\xf\rangle$ & $\langle\pt\rangle$  & \multicolumn{1}{c}{Polarization} \\ 
& $(10 ^{-2})$ & (\gevc) &\\ \hline
Integrated & $-2.7$	&	1.5	&	$-0.005$	$\pm$	0.008 $\pm$	0.006  \\ \hline
\xf ($\times 10^{-2}$)   & & &   \\ 
$[-10.0 , -5.5]$	&	$-6.4$	&	2.8	&	0.027 $\pm$ 0.050 $\pm$ 0.026 \\
$[-5.5 , -3.0]$	&	$-3.8$	&	1.9	&	0.002 $\pm$ 0.015 $\pm$ 0.017 \\
$[-3.0 , -2.0]$	&	$-2.4$	&	1.5	&	$-0.009$ $\pm$ 0.013 $\pm$ 0.011 \\
$[-2.0 , -1.0]$	&	$-1.7$	&	1.3	&	$-0.009$ $\pm$ 0.019 $\pm$ 0.026 \\

 \hline
\pt (\gevc)   & & &  \\ 
$[0.15 , 1.00]$	&	$-2.1$	&	0.8	&	$-0.002$ $\pm$ 0.021 $\pm$ 0.023 \\
$[1.00 , 1.50]$	&	$-2.4$	&	1.2	&	$-0.023$ $\pm$ 0.014 $\pm$ 0.013 \\
$[1.50 , 2.00]$	&	$-2.9$	&	1.7	&	$-0.004$ $\pm$ 0.016 $\pm$ 0.014 \\
$[2.00 , 2.50]$	&	$-3.3$	&	2.2	&	0.010 $\pm$ 0.022 $\pm$ 0.018 \\
$[2.50 , 3.00]$	&	$-3.5$	&	2.7	&	$-0.006$ $\pm$ 0.031 $\pm$ 0.018 \\
$[3.00 , 4.00]$	&	$-3.8$	&	3.4	&	0.082 $\pm$ 0.037 $\pm$ 0.019 \\
$[4.00 , 6.00]$	&	$-4.0$	&	4.6	&	$-0.002$ $\pm$ 0.069 $\pm$ 0.048 \\
\hline
    \end{tabular}
\end{table}

\section{Conclusions}
The $\Lz$ and $\Lbar$ transverse polarization has been measured in $p$Pb collisions at \mbox{$\sqsnn=5.02\tev$} as a function of \pt and \xf. The results are consistent with zero for both $\Lz$ and $\Lbar$ in the forward and backward rapidity regions. This work expands the kinematic range of previous measurements, bridging the gap between midrapidity ATLAS measurements at low \xf~\cite{ATLAS:2014ona} and other fixed-target measurements at high \xf with different collision energies and configurations~\cite{Lundberg:1989hw,Ramberg:1994tk,Fanti:1998px,LHCb-PAPER-2024-009}. The $p$Pb results from LHCb agree with previous measurements, adding new data at different energy and collision configurations within a kinematic range that was previously underexplored. The \pt in this measurement extends to values higher than fixed-target experiments that have observed nonzero transverse $\Lz$ polarization. Compared with earlier observations of significant transverse \Lz polarization in fixed-target data, the results presented in this paper indicate that the polarization effects are small towards low \xf at both positive and negative rapidity. These results contribute to ongoing efforts in understanding transverse spin effects in hadron collisions, providing constraints on the magnitude and kinematic dependence of fragmentation functions in a nuclear environment. In particular, they can offer input and tests for theoretical frameworks based on polarizing TMD FFs and multiparton correlators for the low-\xf region and asymmetric nuclear collisions. 

\section*{Acknowledgements}
%
%
\noindent We express our gratitude to our colleagues in the CERN
accelerator departments for the excellent performance of the LHC. We
thank the technical and administrative staff at the LHCb
institutes.
We acknowledge support from CERN and from the national agencies:
ARC (Australia);
CAPES, CNPq, FAPERJ and FINEP (Brazil); 
MOST and NSFC (China); 
CNRS/IN2P3 (France); 
BMBF, DFG and MPG (Germany); 
INFN (Italy); 
NWO (Netherlands); 
MNiSW and NCN (Poland); 
MCID/IFA (Romania); 
MICIU and AEI (Spain);
SNSF and SER (Switzerland); 
NASU (Ukraine); 
STFC (United Kingdom); 
DOE NP and NSF (USA).
We acknowledge the computing resources that are provided by ARDC (Australia), 
CBPF (Brazil),
CERN, 
IHEP and LZU (China),
IN2P3 (France), 
KIT and DESY (Germany), 
INFN (Italy), 
SURF (Netherlands),
Polish WLCG (Poland),
IFIN-HH (Romania), 
PIC (Spain), CSCS (Switzerland), 
and GridPP (United Kingdom).
We are indebted to the communities behind the multiple open-source
software packages on which we depend.
Individual groups or members have received support from
Key Research Program of Frontier Sciences of CAS, CAS PIFI, CAS CCEPP, 
Fundamental Research Funds for the Central Universities,  and Sci.\ \& Tech.\ Program of Guangzhou (China);
Minciencias (Colombia);
EPLANET, Marie Sk\l{}odowska-Curie Actions, ERC and NextGenerationEU (European Union);
A*MIDEX, ANR, IPhU and Labex P2IO, and R\'{e}gion Auvergne-Rh\^{o}ne-Alpes (France);
Alexander-von-Humboldt Foundation (Germany);
ICSC (Italy); 
Severo Ochoa and Mar\'ia de Maeztu Units of Excellence, GVA, XuntaGal, GENCAT, InTalent-Inditex and Prog.~Atracci\'on Talento CM (Spain);
SRC (Sweden);
the Leverhulme Trust, the Royal Society and UKRI (United Kingdom).


\addcontentsline{toc}{section}{References}
\bibliographystyle{LHCb/LHCb}
\bibliography{main,LHCb/standard,LHCb/LHCb-PAPER,LHCb/LHCb-CONF,LHCb/LHCb-DP,LHCb/LHCb-TDR}

\newpage
\centerline
{\large\bf LHCb collaboration}
\begin
{flushleft}
\small
R.~Aaij$^{38}$\lhcborcid{0000-0003-0533-1952},
A.S.W.~Abdelmotteleb$^{57}$\lhcborcid{0000-0001-7905-0542},
C.~Abellan~Beteta$^{51}$\lhcborcid{0009-0009-0869-6798},
F.~Abudin{\'e}n$^{57}$\lhcborcid{0000-0002-6737-3528},
T.~Ackernley$^{61}$\lhcborcid{0000-0002-5951-3498},
A. A. ~Adefisoye$^{69}$\lhcborcid{0000-0003-2448-1550},
B.~Adeva$^{47}$\lhcborcid{0000-0001-9756-3712},
M.~Adinolfi$^{55}$\lhcborcid{0000-0002-1326-1264},
P.~Adlarson$^{84}$\lhcborcid{0000-0001-6280-3851},
C.~Agapopoulou$^{14}$\lhcborcid{0000-0002-2368-0147}
% ...

{\footnotesize \it

% ...
$^{14}$Universit{\'e} Paris-Saclay, CNRS/IN2P3, IJCLab, Orsay, France\\
% ...
$^{38}$Nikhef National Institute for Subatomic Physics, Amsterdam, Netherlands\\
% ...
$^{47}$Instituto Galego de F{\'\i}sica de Altas Enerx{\'\i}as (IGFAE), Universidade de Santiago de Compostela, Santiago de Compostela, Spain\\
% ...
$^{51}$Physik-Institut, Universit{\"a}t Z{\"u}rich, Z{\"u}rich, Switzerland\\
% ...
$^{55}$H.H. Wills Physics Laboratory, University of Bristol, Bristol, United Kingdom\\
% ...
$^{57}$Department of Physics, University of Warwick, Coventry, United Kingdom\\
% ...
$^{60}$School of Physics and Astronomy, University of Glasgow, Glasgow, United Kingdom\\
$^{61}$Oliver Lodge Laboratory, University of Liverpool, Liverpool, United Kingdom\\
% ...
$^{69}$Syracuse University, Syracuse, NY, United States\\
% ...
$^{84}$Department of Physics and Astronomy, Uppsala University, Uppsala, Sweden, associated to $^{60}$\\
% ...
}
\end{flushleft}

\end{document}